# Diamond anvil cell with boron-doped diamond heater for high-pressure synthesis and in-situ transport measurements

*Ryo Matsumoto[a], Sayaka Yamamoto[b,c], Shintaro Adachi[d], Takeshi Sakai[e], Tetsuo Irifune[e], Yoshihiko Takano[b,c]

[a]International Center for Young Scientists (ICYS),
National Institute for Materials Science, Tsukuba, Ibaraki 305-0047, Japan
[b]International Center for Materials Nanoarchitectonics (MANA),
National Institute for Materials Science, Tsukuba, Ibaraki 305-0047, Japan
[c]Graduate School of Pure and Applied Sciences, University of Tsukuba, 1-1-1 Tennodai, Tsukuba, Ibaraki 305-8577, Japan
[d]Nagamori Institute of Actuators, Kyoto University of Advanced Science, Ukyo-ku, Kyoto 615-8577, Japan
[e]Geodynamics Research Center (GRC), Ehime University, Matsuyama, Ehime 790-8577, Japan
[f]Research Center for Functional Materials,
National Institute for Materials Science, Tsukuba, Ibaraki 305-0047, Japan

*Corresponding author; Email: matsumoto.ryo@nims.go.jp

**Abstract**

Temperature and pressure are essential parameters in the synthesis, evaluation, and application of functional materials. This study proposes the addition of a heating function to a high-pressure diamond anvil cell (DAC) with in-situ measurement probes. The proposed DAC allows for simultaneous control of temperature and pressure within the sample space and can be used to synthesize functional materials under extreme conditions. The various components, namely the heater, thermometer, and measurement probes, were fabricated with a boron-doped diamond epitaxial film and could be used repeatedly. The developed DAC was used to successfully conduct the high-pressure annealing of La(O,F)BiS$_2$ single crystal and the high-pressure synthesis of EuFBiS$_2$ superconductors. The proposed technique shows promise for further exploration of superconductors to broaden the research field.

Superconductors are promising materials for various applications such as energy-saving products, super-fast trains, and high-resolution magnetic sensors. However, superconductivity can only be achieved within limited conditions under a critical temperature ($T_c$) of less than 20 K for NbTi- and Nb$_3$Sn-based materials [1]. Liquid nitrogen is needed, even for high-$T_c$ cuprate cables of Bi$_2$Sr$_2$Ca$_2$Cu$_3$O$_{10+\delta}$ (DI-BSCCO) with a higher $T_c$ of 110 K [2]. This limitation has served as a bottleneck in the widespread use of superconducting products. However, high-$T_c$ superconductivity has been recently reported in some metal hydrides, including H$_3$S [3-5], LaH$_x$ [6,7], YH$_x$ [8], TmH$_x$ [9], SnH$_x$ [10], and carbonaceous sulfur hydride [11], when synthesized within a high-pressure apparatus of diamond anvil cell (DAC). The design of hydrogen-based superconductors remains a challenge because the previously reported materials only exist under high pressures of above 100 GPa. Thus, the development of high-$T_c$ superconductors at lower pressure is important for practical application.

The synthesis of superconducting hydrides is affected by several issues. First, electrical measurement probes are required to observe the superconductivity of the sample under extreme conditions. Specialized DACs with boron-doped diamond (BDD) probes have been developed to function stably under high pressure due to their superior chemical and mechanical hardness [12-14]. These BDD probes can be used for repeated resistivity measurements until the diamond anvil itself is broken. An observation of the superconducting phase with $T_c$ of 25 K on sulfur hydrides at a high pressure above 120 GPa has been successfully demonstrated using the DAC with BDD probes [15].

The second issue is the need for a high-temperature generation method to react the hydrogen and host metal under high pressure. Laser heating is generally used to synthesize superconducting hydrides and achieve high temperatures beyond 5000°C [16-18]. However, temperature control below 1000 °C is difficult due to the large temperature distribution. Consequently, the exact temperatures for the synthesis of superconducting hydrides are still unclear in previous studies [5-11]. Alternatively, resistive heating can be performed using an internal heater placed inside the sample chamber [19]. The method is generally used in a multi-anvil press and provides superior temperature control. However, the fabrication of a resistive heater inside the DAC sample chamber is complex [20]. Further, the resistive heater is destroyed during compression. Therefore, an innovative technique for sample heating is required to fabricate functional materials under high pressure.

This study proposes a high-temperature and high-pressure sample synthesis platform with an in-situ electrical transport measurement function. As shown in the schematic image of Fig. 1 (a), the various components, including transport measurement probes, a resistive heater, and a resistive thermometer, are fabricated on the diamond anvil. The temperature of the sample space is controlled by tuning the input power to the heater on the diamond anvil. In-situ electrical transport measurements of the sample are performed during compression and heating, and X-ray or laser analyses are possible due to the transparency of the diamond anvil. Kinetic analysis of the samples is also available using the in-situ measurement function to monitor the chemical reaction during heating and compression. The measurement probes, resistive heater, and resistive thermometer are fabricated from a BDD film homoepitaxially grown from the diamond anvil surface and could all be used repeatedly. A typical circuit schematic of the proposed DAC with BDD probes, heater, and the thermometer is presented in Fig. 1 (b). The heater is connected to a power supply via a shunt resistor to monitor the circuit current.

The resistances of the sample and thermometer are measured using a standard four-probe method, and the generated temperature is determined based on thermometer resistance using a calibration table.

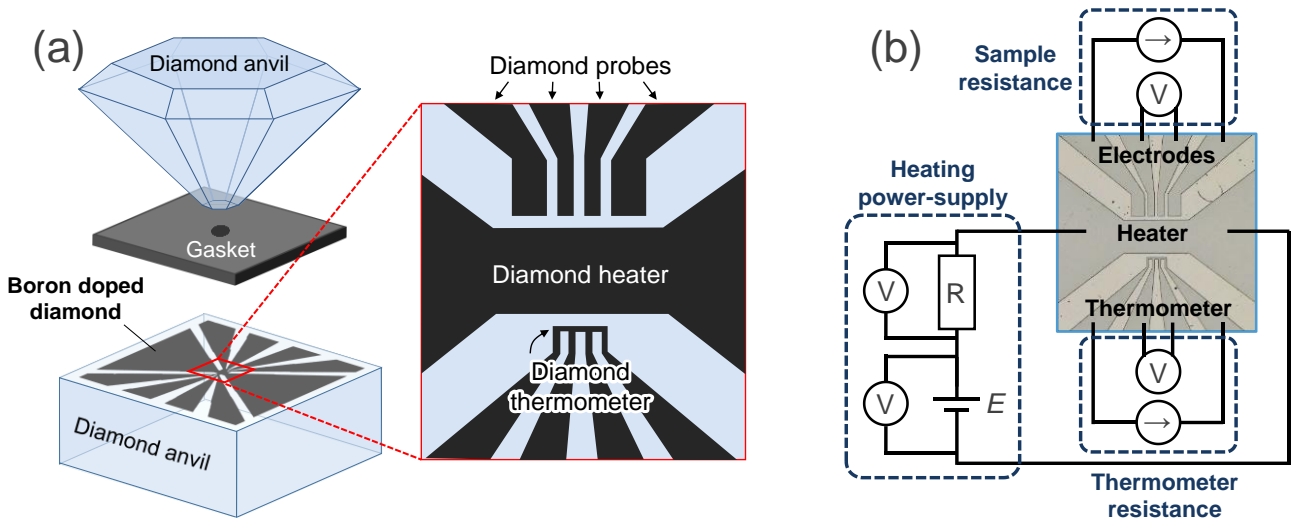

**Figure 1. (a) Schematic illustration of the proposed DAC. (b) Typical circuit schematic for a heating experiment.**

The fabrication of the BDD components was conducted via a combination of electron beam lithography and microwave plasma-assisted chemical vapor deposition (Fig. 1(b)), as previously described in the literature [12-14]. The fabrication approach is applicable to various kinds of the diamond anvils, such as a culet-type of single-crystalline anvils and nano-polycrystalline anvils [21]. The fabricated anvil was subjected to a heating test at ambient pressure. The calibration table to determine the temperature based on thermometer resistance was established by measuring the resistance-temperature (*R-T*) relationship in a tube furnace under $N_2$ gas flow to avoid oxidation of the diamond. The temperature was measured using a thermocouple placed near the diamond anvil. The thermometer exhibited metallic behavior until ~800 K, as shown in the temperature dependence of resistance in Fig. 2 (a). Although the cause of *R-T* curve saturation is currently unknown, it can be prevented by tuning a boron concentration (Fig. S1). Thus, the heavily doped diamond (i.e., metallic diamond) exhibited a plateau within the high-temperature region of the *R-T* curve, while the lightly doped insulating diamond exhibited no saturation. Figure 2 (b) shows the input power dependence of thermometer resistance of the heater anvil. The resistance was successfully converted to generated temperature, as shown in Fig. 2 (c), using the established temperature calibration table. The temperature measurements were limited at 800 K due to saturation of the temperature calibration table, but the anvil emitted a glowing red color at higher input powers.

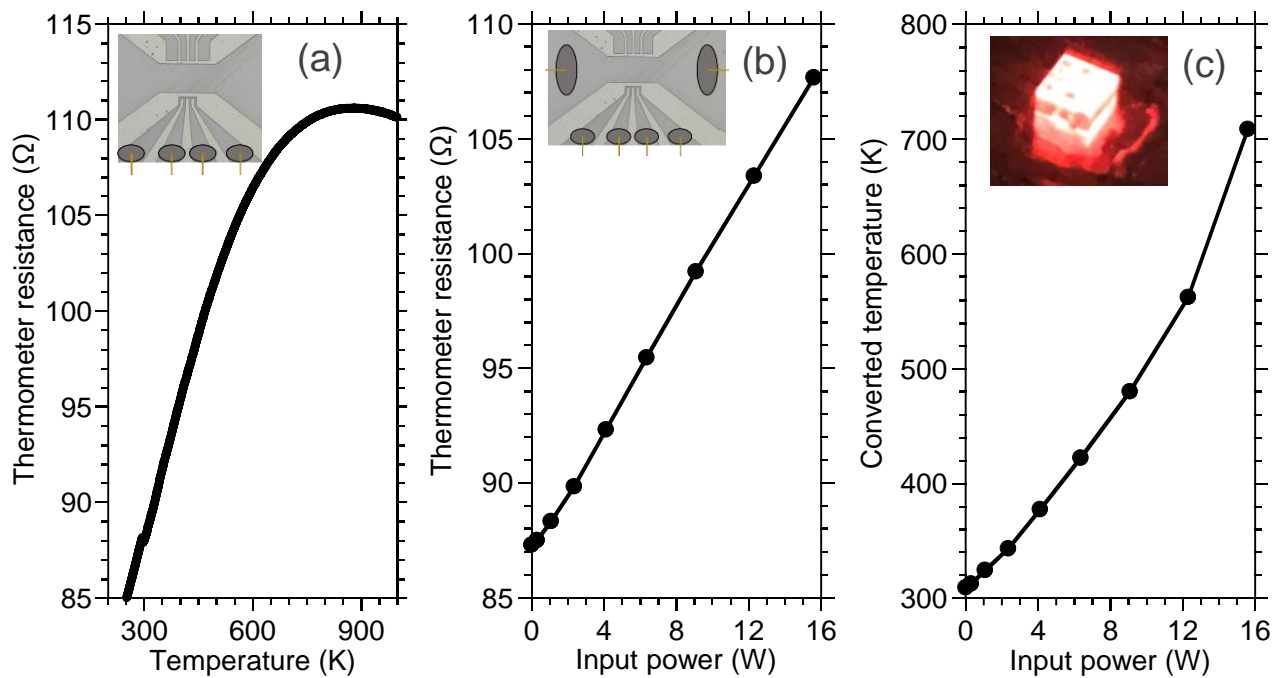

**Figure 2.** Relationship between (a) thermometer resistance and temperature, (b) thermometer resistance and input power, and (c) input power and converted temperature (inset: photograph of the diamond anvil at a high input power above 16 W).

The DAC for the high-pressure heating test was prepared, as shown in a schematic image of Fig. 3 (a). The configuration included a SUS316 stainless steel gasket, cubic boron nitride (cBN) pressure-transmitting medium, and ruby powder manometer, while the diamond anvils were fixed to the backup plates using Ag paste. The material of the backup plate should be carefully selected based on mechanical hardness and thermal conductivity, where candidates include WC, $Si_3N_4$, $ZrO_2$, $B_4C$, and $Al_2Si_4O_{10}(OH)_2$. In this heating test, $Si_3N_4$ was selected for the backup plates. The generated pressure was determined using the ruby fluorescence method [22] at room temperature. The DAC was placed in a glass tube under $N_2$ gas flow to avoid oxidation of the diamond. The radiation temperature was measured using an optical fiber cable to estimate the higher temperature region beyond 800 K, while the surface temperature of the DAC was measured using a thermocouple to evaluate the thermal diffusion from the sample chamber. The heating test was conducted under 3 GPa (Fig. 3 (b)). The temperature of sample space was systematically controlled by changing the input power, where the heating/cooling rates were dependent on the gas flow rate. Using another configuration of DAC with $Al_2Si_4O_{10}(OH)_2$ backup plate, the accuracy of the temperature estimation was evaluated by plotting the input power by the temperatures measured using the diamond thermometer and radiation thermometer (Fig. S2). The temperatures measured by both thermometers were well-correlated within their overlap region. The temperature of the sample space exceeded 1000 K, and the emission of the red color was observed at the maximum input power. The surface temperature of the cell remained close to room temperature, even when the anvil was glowing red. Overall, the heating tests confirmed that the developed DAC could generate high-temperature and high-pressure conditions using the diamond heater and diamond thermometer.

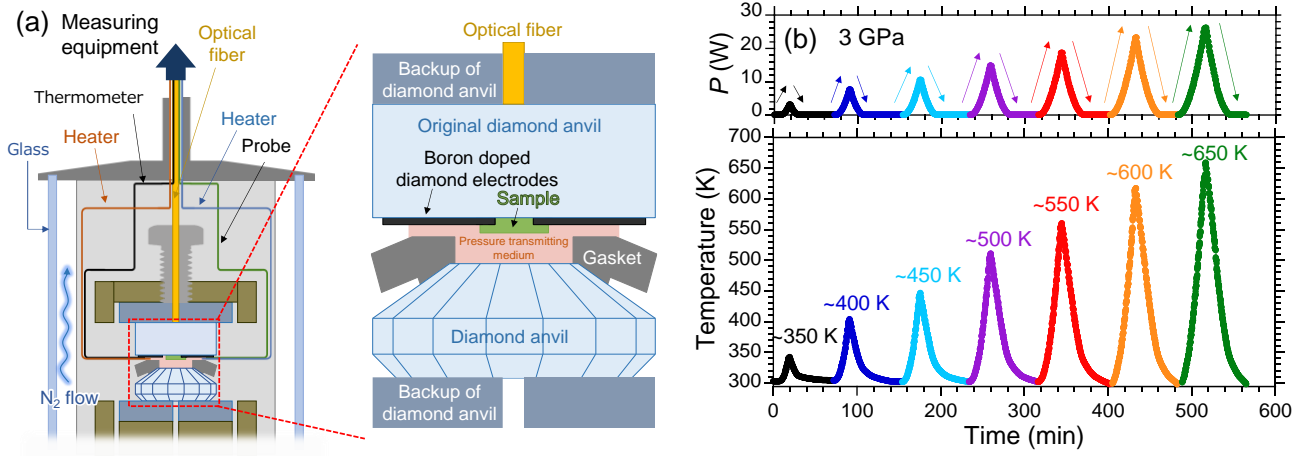

**Figure 3. (a) Schematic illustration of the DAC configuration and measurement system for heating experiments under high pressure. (b) Heating test of the DAC at a pressure of 3 GPa.**

Two different high-pressure and high-temperature experiments were conducted to demonstrate the performance of the developed DAC, namely high-pressure annealing of La(O,F)BiS$_2$ single crystal and high-pressure synthesis of EuFBiS$_2$. The high-pressure application for the layered superconductor La(O,F)BiS$_2$ single-crystal enhances its original $T_c$ of 3 K to 10 K via a structural phase transition only during the compression above 1 GPa [23,24]. This enhanced $T_c$ can be quenched to ambient pressure by high-pressure annealing on the polycrystalline sample [25]. However, the quench of the higher $T_c$ phase has never been reported for a single crystal sample.

Single crystals of La(O,F)BiS$_2$ were grown using an alkali metal flux according to a method similar to previous studies [27,28]. The starting materials of La$_2$S$_3$, Bi$_2$O$_3$, BiF$_3$, Bi$_2$S$_3$, Bi were combined to a nominal composition of LaO$_{0.5}$F$_{0.5}$BiS$_2$, while the flux was prepared by mixing CsCl and KCl in a molar ratio of 5:3. The mixture was sealed in an evacuated quartz tube and annealed according to a previously reported temperature sequence to obtain plate-like single crystals. The details of sample characterization have been previously reported [29]. A cleaved crystal was placed onto the diamond anvil (Fig. 4 (a)), and the DAC was assembled using Al$_2$Si$_4$O$_{10}$(OH)$_2$ as the backup plate material for the diamond anvils.

Figure 4 (b) shows the temperature dependence of the resistance on the La(O,F)BiS$_2$ single crystal under 0.7 GPa. The temperature control below 300 K was performed using a physical property measurement system (PPMS/Quantum Design). The sample resistance gradually decreased with increasing temperature up to 1100 K, namely a semiconducting behavior. The tendency changed to a metallic behavior during the cooling process, which suggested that chemical reactions occurred within the sample at high temperature and pressure. The pressure naturally decreased to ambient pressure as the sample cooled, possibly due to thermal softening of the DAC at high-temperature region. The superconducting properties of La(O,F)BiS$_2$ were compared before and after the annealing (Fig. 4 (c)). The as-grown sample was used as a reference. The $T_c$ was enhanced from 3 K to 8 K by applying pressure before annealing, and the enhanced $T_c$ was maintained even at ambient pressure after annealing. This is the first report of quenching of the high-$T_c$ phase of the La(O,F)BiS$_2$ single crystal via high-pressure annealing. Further studies are recommended to precisely determine the pressure required at low temperatures. The used anvil was acid-cleaned by a mixture of $HNO_3$ and $H_2SO_4$ after the heating experiment (Fig. S3). The electrodes, heater, and thermometer composed of BDD were not degraded after the high-temperature and high-pressure experiment, thereby demonstrating the good stability and reusability of the developed DAC under extreme conditions.

The metallization during high-pressure annealing is considered a key factor in understanding the unique feature on $La(O,F)BiS_2$. Figure 4 (d) shows a repeated measurement of $R$-$T$ property of $La(O,F)BiS_2$ single crystal under 3 GPa using different DAC configulation with the backup plate of $Si_3N_4$ based on the temperature sequence of Fig. 3 (b). First, the temperature was increased up to 350 K and decreased to room temperature, and then, the sample resistance was returned to the original one. By applying the temperature above 400 K, the sample resistance showed an irreversible change. The temperature dependence of the resistance changed from semiconducting to metallic after experiencing 500 K annealing. According to the literature [30], as-synthesized $La(O,F)BiS_2$ exhibits a semiconducting nature and filamentary superconductivity, as seen in Fig. 4 (c), because of the instability of local crystal structure. The application of high pressure stabilizes the structure and induces bulk superconductivity. The observed metallization possibly relates to the appearance of the bulk superconductivity in $La(O,F)BiS_2$.

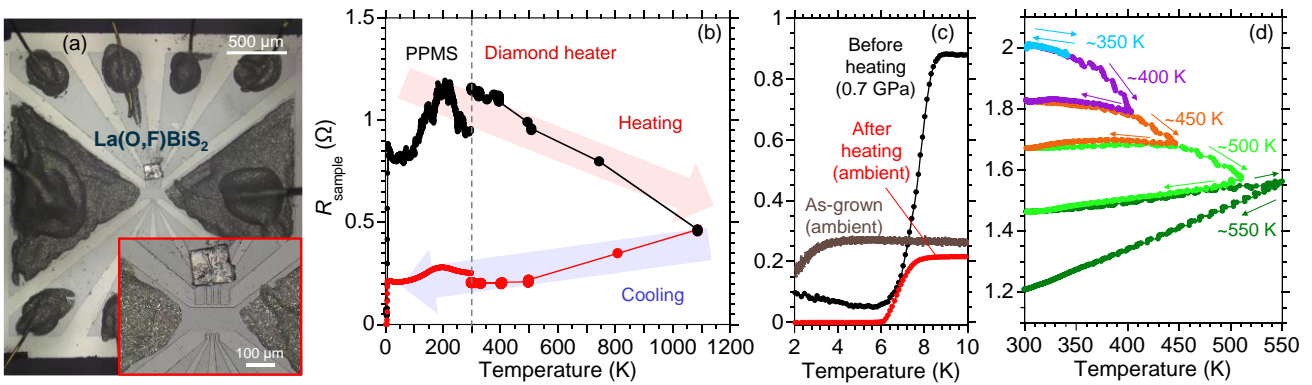

**Figure 4. High-temperature and high-pressure annealing for $La(O,F)BiS_2$ single crystal. (a) Optical image of the diamond anvil. (b) Temperature dependence of the resistance up to 1100 K under 0.7 GPa. (c) Superconducting properties before and after the high-pressure annealing. (d) Repeated measurement of R-T property under 3 GPa.**

The high-pressure synthesis experiment was conducted for a layered superconductor $EuFBiS_2$. According to the previous study [26], $EuFBiS_2$ exhibits pressure-induced superconductivity with a $T_c$ of 5 K at 0.7 GPa. The $T_c$ can be enhanced to 8.5 K under higher pressure. However, the effects of high-temperature and high-pressure synthesis on the superconducting properties of $EuFBiS_2$ have not yet been reported. The temperature and pressure range for the high-pressure synthesis of $EuFBiS_2$ is suitable for the demonstration of the developed DAC.

Starting materials of $EuS$, $Bi_2S_3$, and $BiF_3$ were mixed in a nominal composition. The mixture was placed onto the diamond anvil, and the DAC was prepared using $ZrO_2$ as the backup plate material. The temperature was plotted against the resistance of the starting material mixture under 3.1 GPa, as shown in Fig. 5 (a). Before the heating, the sample exhibited insulating behavior without superconductivity. The temperature was increased up to 900 K and held for 3 min during the heating process, after which the sample was quenched to room temperature. The pressure was naturally decreased by 2.1 GPa during the annealing. The heating induced a drastic reduction in sample resistance at 300 K of more than three orders of magnitude. Although the resistance behavior was still insulating, possibly due to insufficient reaction time, d$R$/d$T$ during cooling was lower than that before the annealing. Moreover, the sample resistance exhibited a sudden drop in the low-temperature region.

A magnified view of the $R$-$T$ curve of the sample under a magnetic field after the heating revealed that the sample resistance started to decrease from 8.6 K (Fig. 5(b)). This drop was gradually suppressed as the magnetic field was increased up to 7 T, suggesting an emergence of superconductivity. The inset of Fig. 5(b) shows a temperature dependence of the critical fields ($B_{c2}$) of the observed superconductivity. The measured $B_{c2}$ is drastically higher than that reported in previous results without heating [26]. Similar anomalous $B_{c2}$ has been reported in the high-pressure phase of La(O,F)BiS$_2$ [31], suggesting common features in the BiS$_2$-based superconductors. The observed $T_c$ value was similar to previous reports [26], thereby confirming the successful high-pressure synthesis of EuFBiS$_2$ using the heater DAC.

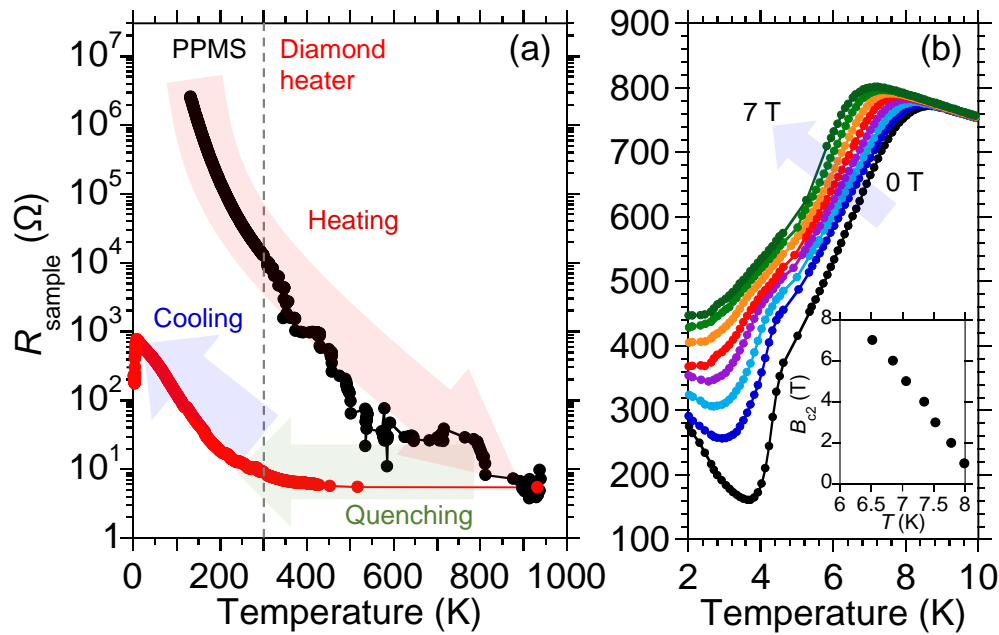

**Figure 5.** (a) Relationship between temperature and resistance on the high-pressure synthesis for EuFBiS$_2$. (b) Magnified view of the $R$-$T$ curve after heating under a magnetic field of 0 to 7 T. The inset shows a temperature dependence of $B_{c2}$.

In conclusion, a DAC with various BDD components, including the heater, thermometer, and probes, was successfully developed to control temperature and pressure simultaneously. The proposed method allowed for high-pressure synthesis and in-situ measurements of electrical transport properties. Further, the reusability of the BDD components was demonstrated after heating tests under pressure. The developed DAC was successfully used to conduct the high-pressure annealing of a La(O,F)BiS$_2$ single crystal and the high-pressure synthesis of EuFBiS$_2$. Higher temperature and pressure generation are expected as the DAC is developed further, promising to synthesize high-$T_c$ superconductors.

**Acknowledgment**

The authors thank Dr. M. Imai, Dr. K. Tsuchiya, Dr. T. Taniguchi, and Dr. T. D. Yamamoto from NIMS for their discussion and valued input. This work was partly supported by JST-Mirai Program Grant Number JPMJMI17A2, JSPS KAKENHI Grant Number JP19H02177, 20H05644, and 20K22420. The fabrication process of diamond electrodes was partially supported by the NIMS Nanofabrication Platform in the Nanotechnology Platform Project sponsored by the Ministry of Education, Culture, Sports, Science and Technology (MEXT), Japan. The high-pressure experiments were supported by the Visiting Researcher's Program of the GRC. The nano-polycrystalline diamond was synthesized by Toru Shinmei of the GRC. RM would like to acknowledge the ICYS Research Fellowship, NIMS, Japan.

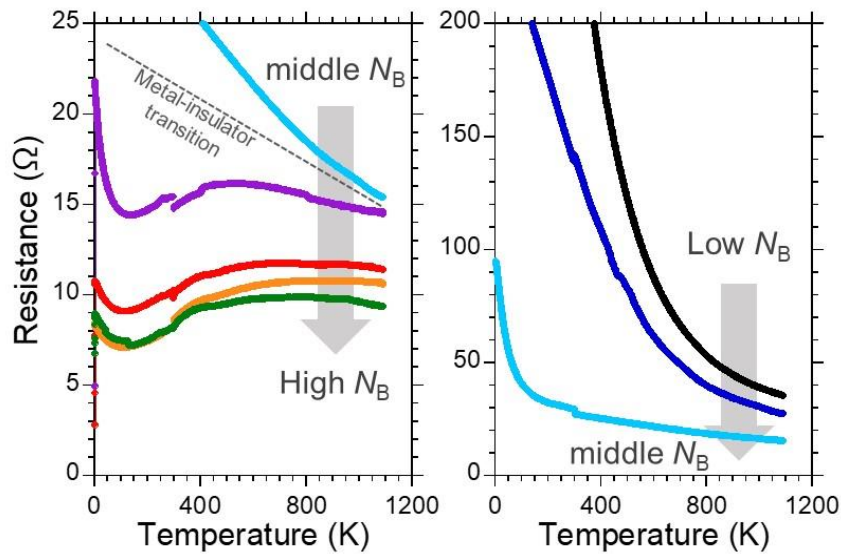

**Figure S1.** Relationship between temperature and resistance of the boron-doped diamond film at various boron-concentrations ($N_B$).

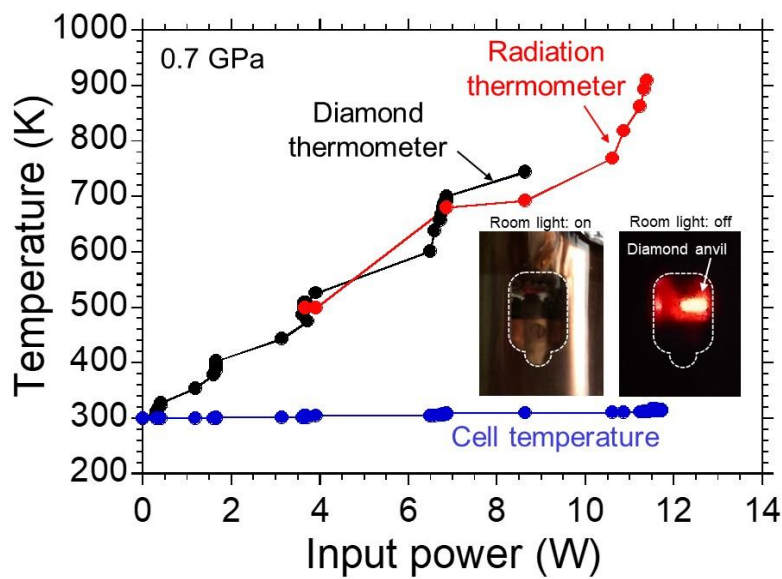

**Figure S2.** Relationship between input power and temperature measurements using the diamond thermometer (black) and radiation thermometer (red).

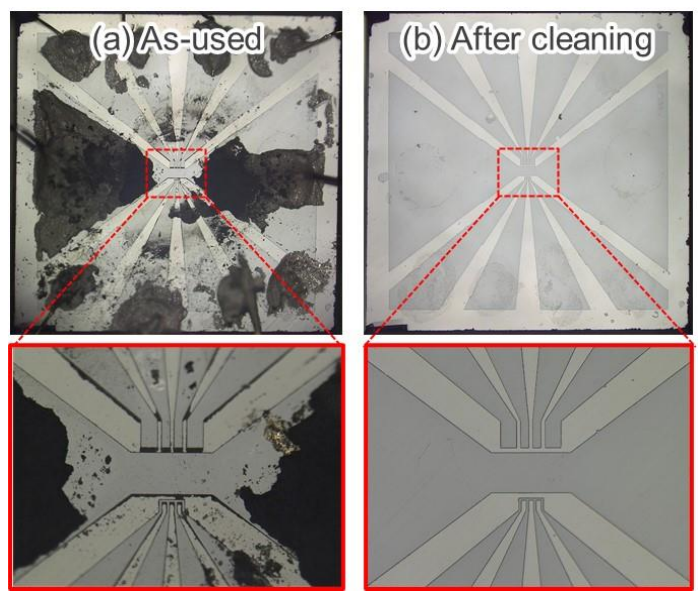

**Figure S3.** Optical image of the (a) used anvil after the heating experiment and (b) after acid-cleaning using mixture of $HNO_3$ and $H_2SO_4$.